\documentstyle[12pt,twoside]{article}
\begin{document}
\begin{center}

{\Large\bf A remark on Brans-Dicke cosmological\\[5pt]
dust solutions with negative $\omega$\\[5pt]}
\medskip
 
{\bf A.B. Batista\footnote{e-mail brasil@cce.ufes.br},
J. C. Fabris\footnote{e-mail: fabris@cce.ufes.br} and
R. de S\'a Ribeiro\footnote{e-mail: ribeiro@cce.ufes.br}} \medskip

Departamento de F\'{\i}sica, Universidade Federal do Esp\'{\i}rito Santo, 
29060-900, Vit\'oria, Esp\'{\i}rito Santo, Brazil
\medskip

\end{center}
 
\begin{abstract}
Analysing the Brans-Dicke solutions for the dust phase, we show that,
for negative values of $\omega$, they contain
scenarios that display an initial subluminal expansion followed by an
inflationary
phase. We discuss these solutions with respect to the results of the
observation of
high redshif supernova as well as the age problem and structure formation.
We stablish possible connections of these solutions with those emerging
from
string effective models.
\vspace{0.7cm}

PACS number(s): 04.20.Cv., 04.20.Me
\end{abstract}
The recent results coming from the high redshift supernova observations
indicate that the Universe
today may be in an accelerated phase \cite{riess,perlmu1}. This suggests
that the energy content of the Universe may be
dominated today by a fluid with negative pressure. The cosmological
constant is a natural candidate for
this dark energy. The cosmological constant may be represented by a
perfect fluid with an equation
of state $p = - \rho$. However, there is a controverse about the effective
equation of state of the
exotic fluid responsable for the acceleration of the expansion. In
\cite{wang} it is claimed that this exotic fluid may be represented by an
equation
$p = \alpha\rho$, with $-1 \leq \alpha \leq -0.6$, while in \cite{perlmu2}
it is found the limits
$-1 \leq \alpha \leq -0.8$. The confirmation of the preliminary results of
the high redshif supernova
program will bring new important questions about the nature of the dark
matter in the Universe.
\par
To take into account the possibility of an inflationary phase in the
Universe today, it has been proposed
recently
a model where a scalar field is minimally coupled to gravity containing a
suitable potential
\cite{caldwell,martin}.
The shape of the potential term is such that the effective equation of
state evolves from 
$\alpha = \frac{1}{3}$, characteristic of a radiative phase, to a negative
value near $-1$.
This model has been called {\it quintessence}. In this letter we would
like to call attention to the fact
that the prototype of a scalar-tensor theory, with a non-minimally coupled
scalar field and
no potential, the Brans-Dicke theory, has a class of solutions for the
dust equation of
state $p = 0$
exhibiting a non-inflationary initial regime and an inflationary phase in
the asymptotic limit.
This class of asymptotic dust inflationary solutions can be consistent, at
least from
the kinetic point of view, with the recent results of an accelerating
Universe. We stress that some studies of quintessence model have been
carried out in the context of
Brans-Dicke theory, but employing a sef-interacting scalar field
\cite{bartolo,bertolami}, what is not the case analyzed here.
\par
In fact, let us consider the Brans-Dicke Lagrangian,
\begin{equation}
L = \sqrt{-g}\biggr(\phi R -
\omega\frac{\phi_{;\rho}\phi^{;\rho}}{\phi}\biggl) \quad .
\end{equation}
The parameter $\omega$ defines the coupling of the scalar field to
gravity.
Considering a Friedmann-Robertson-Walker flat Universe, this Lagrangian
results in
the following equations of motion:
\begin{eqnarray}
\label{em1}
3\biggr(\frac{\dot a}{a}\biggl)^2 &=& \frac{8\pi\rho}{\phi} +
\frac{\omega}{2}\biggr(\frac{\dot\phi}{\phi}\biggl)
- 3\frac{\dot a}{a}\frac{\dot\phi}{\phi} \quad ,\\
\label{em2}
\ddot\phi + 3\frac{\dot a}{a}\frac{\dot\phi}{\phi} &=& \frac{8\pi\rho}{3 +
2\omega} \quad ,\\
\label{em3}
\dot\rho + 3\frac{\dot a}{a}\rho &=& 0 \quad .
\end{eqnarray}
The last equation leads to the $\rho = \rho_0a^{-3}$. Inserting this
relation in (\ref{em2}), we find the first integral,
$\dot\phi = \frac{8\pi\rho t}{3 + 2\omega}C$, where $C$ is a constant.
\par
Following \cite{weinberg}, we can define an auxiliary function $u$,
satisfying the relation
\begin{equation}
\label{rel}
\frac{\dot u}{u} = -3\frac{\dot a}{a} + \frac{2}{t} - \frac{u}{t}
\end{equation}
which, in view of the equations of motion, results in the following
integral relation
\begin{equation}
\label{int}
\int\frac{du}{u[u + 4 \pm a\sqrt{u^2 + 4u}]} = 2\ln (t - t_c)\quad,
\end{equation}
where $a = 3\sqrt{1 + \frac{2}{3}\omega}$. This integral relation has
three critical
points $u = 0$, $u = - 4$ and $u = \frac{2}{4 + 3\omega}$. The first one
gives non physical
results, while the second one leads to $a \propto t^2$ for any value of
$\omega$. However,
the energy density of ordinary matter associated to this solution is
negative.
\par
The third critical point of (\ref{int}) leads to a particular solution in
terms of power law function:
\begin{equation}
\label{ps}
a = a_0t^\frac{2 + 2\omega}{4 + 3\omega} \quad , \quad \phi =
\phi_0t^\frac{2}{4 + 3\omega} \quad .
\end{equation}
A more general solution may be obtained through integration of
(\ref{int}).
This has been donne in \cite{gurevich}, resulting in the general flat
solution
\begin{eqnarray}
\label{g1}
a &=& a_0(t - t_-)^\frac{1 + \omega \pm \sqrt{1 + \frac{2}{3}\omega}}{4 +
3\omega}(t - t_+)^\frac{1 + \omega
\mp \sqrt{1 + \frac{2}{3}\omega}}{4 + 3\omega} \quad , \\
\label{g2}
\phi &=& \phi_0(t - t_-)^\frac{1 \mp 3\sqrt{1 + \frac{2}{3}\omega}}{4 +
3\omega}
(t - t_+)^\frac{1 \pm 3\sqrt{1 + \frac{2}{3}\omega}}{4 + 3\omega} \quad .
\end{eqnarray}
Since $t_+$ and $t_-$ are constants such that $t_+ > t_-$ , $t = t_+$ may
be identified with
the initial time.
\par
Let us first inspect the solutions (\ref{g1},\ref{g2}) in more details.
They have two asymptotic regimes. Near the initial singularity, we find
\begin{eqnarray}
\label{i1}
a &=& a_0t^\frac{1 + \omega \pm \sqrt{1 + \frac{2}{3}\omega}}{4 +
3\omega}\quad , \\
\label{i2}
\phi &=& \phi_0t^\frac{1 \mp 3\sqrt{1 + \frac{2}{3}\omega}}{4 + 3\omega}
\quad ,
\end{eqnarray}
where we have redefined $t - t_+ \rightarrow t$.
This is equivalent to the Brans-Dicke vacuum solution. This curious
equivalence is easily understood
if we look at equations (\ref{em1},\ref{em2}): inserting in
(\ref{em1},\ref{em2}) the expressions (\ref{i1},\ref{i2}),
we see that the term $\frac{\rho}{\phi}$ behaves as $\frac{1}{t}$, while
the
terms $(\frac{\dot a}{a})^2, (\frac{\dot\phi}{\phi})^2,
\frac{\dot a}{a}\frac{\dot\phi}{\phi}$ behave as $\frac{1}{t^2}$; for
small $t$, the latter
terms dominate over the former one, the energy of the scalar field
dominating over
the energy of matter, and the vacuum solution is a good approximation.
For large values of $t$, solutions (\ref{g1},\ref{g2}) goes to (\ref{ps}).
\par
In general, the solutions (\ref{g1},\ref{g2}) represent a subluminal
expansion, leading to a yonguer
Universe than in the standard model based in the General Relativity
theory. In the limit $\omega \rightarrow \infty$
they coincide with the dust standard model. The inverse of $\phi$ is
linked with the gravitational coupling
through the relation $G = \frac{4 + 2\omega}{3 + 2\omega}\frac{1}{\phi}$.
Hence, in general,
solutions (\ref{g1},\ref{g2}) predict a
decreasing gravitational coupling, what is reasonable since the Universe
is going asymptotically to a flat
geometry.
\par
However, all the description made above is valid for a positive $\omega$.
More precisely, it is valid for
$\omega > - 1$.
We intend to analyze here what happens if we allow $\omega$ to be
negative. We will first make some general considerations.
If we have $-\frac{4}{3} < \omega < - 1$, the cosmic time must vary as $-
\infty < t \leq 0$,
in order to have an expanding Universe. It represents what is generally
called a pole-law inflation.
If $- \frac{3}{2} < \omega < - \frac{4}{3}$, we have also inflation, but
with $0 \leq t < \infty$.
From these considerations, we have three special values of $\omega$:
$\omega = -1, - \frac{4}{3}$ and $- \frac{3}{2}$.
The case $\omega = - \frac{3}{2}$ represents a breakdown of the theory,
since the Brans-Dicke Lagrangian
may be recast in a form characteristic of a scalar field conformally
coupled to gravity; $\omega = - 1$
leads to a constant scale factor with a varying cosmological constant
(notice that it represents
also the string Lagrangian in presence of ordinary matter); $\omega = -
\frac{4}{3}$ is
a special case which we will discuss latter.
\par
Let us consider the solutions (\ref{g1},\ref{g2}) in the range $-
\frac{3}{2} < \omega < - \frac{4}{3}$.
Initially, for small $t$, we have two branches connected with the signs of
the exponent of 
(\ref{i1},\ref{i2}). The upper sign represents a subluminal expansion,
while the
lower sign a superluminal expansion. In
the limit of large values of $t$, we obtain
the solutions (\ref{ps}) which describe, in the same range of values of
$\omega$, an inflationary Universe.
Hence, it is possible to have the following scenario: the Universe enters
in the
dust phase with a subluminal expansion, evolving latter to a superluminal
expansion.
For example, if we choose $\omega = - 1.4$ and the upper sign in
(\ref{i1},\ref{i2}),
the scale factor behaves initially as $a \propto t^{0.7}$ and as $a
\propto t^4$ in
a latter phase.
\par
Concerning the observational limits for the effective equation of state,
they are obtained essentially by
inspecting the kinematical properties of the scale factor. In the standard
model a barotropic equation
of state leads to the solution
\begin{equation}
\label{sm}
a = a_0t^\frac{2}{3(1 + \alpha)} \quad .
\end{equation}
We will look for values of $\omega$ in (\ref{ps}) that give the same
behaviour as
(\ref{sm}) for a given value of $\alpha$.
It has been argued that the most favoured value for the effective equation
of state today
seems to be $\alpha \sim - 0.77$\cite{efsta}.
This can also be obtained from the matter dominated Brans-Dicke solutions,
in the limit
of large $t$, if
$\omega \sim -1.4329$. Notice that $\alpha = - 1$ corresponds to $\omega =
- \frac{4}{3}$.
The other limit of validity of the Brans-Dicke cosmological solution,
$\omega = - \frac{3}{2}$,
corresponds to $\alpha = - \frac{2}{3}$ (a fluid of cosmic wall). Hence,
$- \frac{3}{2} <
\omega < - \frac{4}{3}$ corresponds, from the kinetic point of view, to $-
1 < \alpha < - \frac{2}{3}$
in the Standard Model.
\par
We go back now to the case $\omega = - \frac{4}{3}$. It can be seen from
the solutions (\ref{ps},\ref{i1},\ref{i2})
that this value for $\omega$ is not allowed. In the case of the "vacuum"
solution, it can be shown that
the power law solution is not possible for this value of $\omega$;
however, we can find exponential solutions:
\begin{equation}
a \propto e^{t} \quad , \quad \phi \propto e^{-3t} \quad .
\end{equation}
For the deep material phase ($t \rightarrow \infty$), we can integrate
exactly (\ref{int})
for $\omega = - \frac{4}{3}$, obtaining $a \propto te^\frac{t}{3}$.
However,
for this solution the energy density is negative. Hence, there is a
pathology for the dust solutions for this
specific value of $\omega$.
\par
The scenario described here must be tested against two important points.
The first one concerns the evolution of density perturbations.
The problem of density perturbations in Brans-Dicke theory has been
studied in \cite{fabris,sergio}.
It can be verified that the inflationary solutions in the dust Universe
described here leads,
in the large wavelength limit, only to decreasing modes. It must be noted
that this is
a generally feature of density perturbations in an inflationary phase
\cite{jerome}.
However, for the upper sign in (\ref{i1},\ref{i2}), the initial behaviour
is subluminal
and very similar to the standard one based in the General Relativity
theory.
Consequently, there is initially growing modes for density perturbations,
and we can expect the formation
of galaxies in the same period as in the standard model.
Moreover, even for large values of $t$,
we must remark that, in opposition to the standard models with negative
pressure,
there is no instability in the small wavelength limit
\cite{fabris,sergio}.
In the quintessence model,
there is only decreasing modes in the large wavelength limit, but the
effective equation of
state at small scale becomes positive, and the perturbations in this limit
oscillate
as an accoustic mode, and no instability is present also. 
\par
Since for the range of $\omega$ defined above the expansion of the
Universe is always
faster than in the standard model, the Universe in this case is older than
in the standard model.
This is a nice feature not only concerning the age problem, but also with
respect to the
structure formation problem.
\par
Secondly, there is the question of the local tests.
This leads to the problem of spherical symmetric solutions for
the Brans-Dicke theory.
It has been shown \cite{lousto,kirill}
that there exist black holes in the Brans-Dicke theory only for negative
values of $\omega$.
In a matter of fact, black holes can exist only for $\omega < -
\frac{4}{3}$. These black
holes are very special since their Hawking temperature is zero and the
area of the horizon
surface is infinite. The local test can be satisfied if $\vert\omega\vert
\geq 500$,
$\omega$ being positive or negative \cite{lousto}.
This fact leads us to consider two main possibility to reconcile local
test with
the considerations made above: localy the scalar field
could be essentially constant, what may suggest a scale dependent
gravitational coupling,
as it has already been evoked in the literature \cite{kim}; or we may
consider a variable
$\omega$, like a non-linear sigma model. Notice that it is possible to
have, when
$\omega =$ constant,
cosmological solutions for $\omega < - \frac{3}{2}$ \cite{gurevich}, but
it predicts, for the
dust phase, a bouncing Universe; moreover, when transposed, through a
conformal
transformation, to the Einstein frame, such case exhibits negative kinetic
energy,
and we must be cautious about its stability.
\par
We remember that the string theory predicts a low energy effective model
with $\omega = - 1$. However, in the case of d-branes string model, the
value
of $\omega$ is given by \cite{Duff}
\begin{equation}
\omega = - \frac{(D - 1)(d - 2) - d^2}{(D - 2)(d - 2) - d^2}
\end{equation}
where $D$ and $d - 1$ are the dimension of space-time and of the brane,
respectivelly.
For example,  for $D = 4$, a 0-brane and 1-brane give $\omega = -
\frac{4}{3}$ and
$\omega = - 1$ respectivelly. These values are not very satisfactory with
respect to
the range of $\omega$ considered above. But, it suggest that it might be
possible that some specific configuration of
the string effective model may lead to those values of $\omega$.

\end{document}